\documentclass{ws-ijmpcs}
\usepackage{url,overcite}

\def\trimmarks{}

\usepackage{multicol,psfrag,cite,cancel}
\usepackage[svgnames]{xcolor}

\newcommand{\be}{\begin{equation}}
\newcommand{\ee}{\end{equation}}
\newcommand{\bea}{\begin{eqnarray}}
\newcommand{\eea}{\end{eqnarray}}

\begin{document}

\markboth{Y.-T. Chien}
{Resummation of Jet Shapes and Extracting QGP Properties}

%
\catchline{}{}{}{}{}
%

\title{Resummation of Jet Shapes and Extracting Properties of the Quark-Gluon Plasma}

\author{Yang-Ting Chien}

\address{Theoretical Division, T-2,
Los Alamos National Laboratory, MS B283\\
Los Alamos, NM 87545, USA\\
ytchien@lanl.gov}

\maketitle

\begin{history}
\received{Day Month Year}
\revised{Day Month Year}
\published{Day Month Year}
\end{history}

\begin{abstract}
Understanding the properties of the quark-gluon plasma (QGP) that is produced in ultra-relativistic nucleus-nucleus collisions has been one of the top priorities of the heavy ion program at the LHC. Energetic jets are produced and subsequently quenched in the collisions. Such jet quenching phenomena provide promising tools to probe the medium properties by studying the modification of jets due to the medium interactions. Significant modifications of jet shapes 
have been measured. In this talk we focus on the calculation of jet shapes in both proton-proton and lead-lead collisions using soft-collinear effective theory (SCET), with Glauber gluon interactions in the medium. Large logarithms in jet shapes are resummed at next-to-leading logarithmic (NLL) accuracy by the renormalization-group evolution between hierarchical jet scales. The medium interactions contribute as power corrections, and we calculate the modification of jet shapes at leading order in opacity with the static QGP model. Preliminary results are presented with good agreement with the recent CMS jet shape measurements.

\end{abstract}

\ccode{PACS numbers: 12.38.Cy, 12.39.St, 13.87.-a, 24.85.+p, 25.75.Bh}

\section{Introduction}

Jets are copiously produced at energetic colliders, and they are manifestation of the partons from the underlying hard scattering processes. Since quantum chromodynamics (QCD) was formulated, jet (substructure) observables played an important role in testing the SU(3) gauge theory structure of perturbative QCD \cite{Farhi:1977sg, Georgi:1977sf, PhysRevLett.41.1581, PhysRevLett.41.1585, PhysRevD.19.2018, Heister:2003aj, Abdallah:2003xz, Achard:2004sv, Abbiendi:2004qz}. The increased theoretical understanding of QCD allowed accurate event shape calculations in $e^+e^-$ collisions and the extraction of the strong coupling constant below percent level uncertainty \cite{Becher:2008cf, Chien:2010kc, Davison:2008vx, Abbate:2010xh}. At hadron colliders, useful techniques have also been developed to look into the substructure of jets and deal with the backgrounds from beam remnant, underlying events and pileup. For a summary of the state-of-the-art studies of jets mostly in proton collisions, see \cite{Altheimer:2013yza}.

In ultra-relativistic heavy ion collisions at RHIC and the LHC, energetic jets are also produced everywhere. The studies of jets in heavy ion collisions have thus drawn considerable attention in the high-energy nuclear physics community. It has been observed that jets are significantly modified or quenched when compared with the jets produced in proton collisions. Such phenomena of jet quenching gives strong evidence of the creation of the hot, dense medium in the collisions \cite{Aad:2012vca,Aad:2010bu,Chatrchyan:2011sx}. The novel medium has been commonly referred to as the quark-gluon plasma (QGP).

One of the top priorities of the heavy ion program at the LHC is to understand the properties of the QGP and how they evolve with energy, and jets provide a unique hard probe of the medium (see Ivan's talk \cite{Vitev:2014qcd}). There has been much effort on the understanding of the suppression in the inclusive jet and leading hadron production cross sections, as well as the modification of jet kinematics. On the other hand, jet substructure observables are sensitive to the details of medium interactions therefore should be ideal for the precise extractions of the medium properties.

The jet shape \cite{Ellis:1992qq} is one of the classic jet substructure observables which probes the transverse energy distribution inside a jet. In heavy ion collisions, the in-medium parton shower is modified and the energy inside jets is redistributed. This leads to the modification of jet shapes which contains useful information about the QGP. In this talk we will present the calculations of jet shapes in both proton-proton \cite{Chien:2014nsa} and lead-lead 
collisions using soft-collinear effective theory (SCET) \cite{Bauer:2000ew, Bauer:2000yr, Bauer:2001ct, Bauer:2001yt, Bauer:2002nz}, with Glauber gluon interactions in the medium \cite{Idilbi:2008vm, Ovanesyan:2011xy, Ovanesyan:2011kn, Fickinger:2013xwa}. As a first attempt we use the static QGP model with phenomenological parameters to describe the medium properties. The calculations with different QGP models are work in progress and hopefully will allow us to pin down the dynamics of QGP.

The rest of the talk is organized as follows. We will introduce the jet shape and discuss its factorization theorem in SCET. Thanks to Chris's talk \cite{Lee:2014xit} on the overview of SCET, we will briefly go through the relevant SCET machinery here. Large logarithms appear in the perturbative calculation of jet shapes which need to be resummed. We discuss how resummation is performed at next-to-leading logarithmic (NLL) accuracy using the renormalization-group techniques. In heavy ion collisions SCET is extended to include Glauber gluon interactions which contribute as power corrections to jet shapes. We use the recently calculated medium-induced splitting functions \cite{Ovanesyan:2011kn, Fickinger:2013xwa} in the jet shape calculation and compare the preliminary results with the recent CMS measurements \cite{Chatrchyan:2013kwa,Kurt:2014mca} with good agreement.

\section{The Jet Shape}
\label{sec:obs}
The integral jet shape $\Psi(r)$ \cite{Ellis:1992qq} of a jet of size $R$ is defined by the fraction of the transverse energy within a distance $r$ from the jet axis $\hat n$,
\be
    \Psi(r)=
    \frac{E_T({\rm inside}~r)}{E_T({\rm inside}~R)}\;.
\ee
After averaging over all jets, the integral jet shape becomes a function of $r$ which describes the average energy distribution inside jets. When $r$ is parametrically smaller than $R$, the perturbative calculation of the jet shape suffers from the presence of large logarithms of the form $\ln r/R$ which need to be resummed. Resummation plays an important role in phenomenology, and we will demonstrate how we achieve it using effective field theory techniques. In fact, the jet shape was resummed using the modified leading logarithmic approximation quite a long time ago \cite{Seymour:1997kj}. The contributions from initial state radiation and non-perturbative effects were also included, which are nevertheless power suppressed and neglected in our framework. See \cite{Li:2011hy, Li:2012bw} for another resummation framework using perturbative QCD. Aside from being a promising probe of QGP properties, the jet shape is actually a powerful jet substructure observable in quark-gluon discrimination \cite{Gallicchio:2011xq,Gallicchio:2012ez} and has a lot of applications in the standard model physics studies and new physics searches.

\section{The Factorization Theorem of the Jet Shape in SCET}
\label{sec:fac}

\begin{figure}
\center
\includegraphics[width=0.28\linewidth]{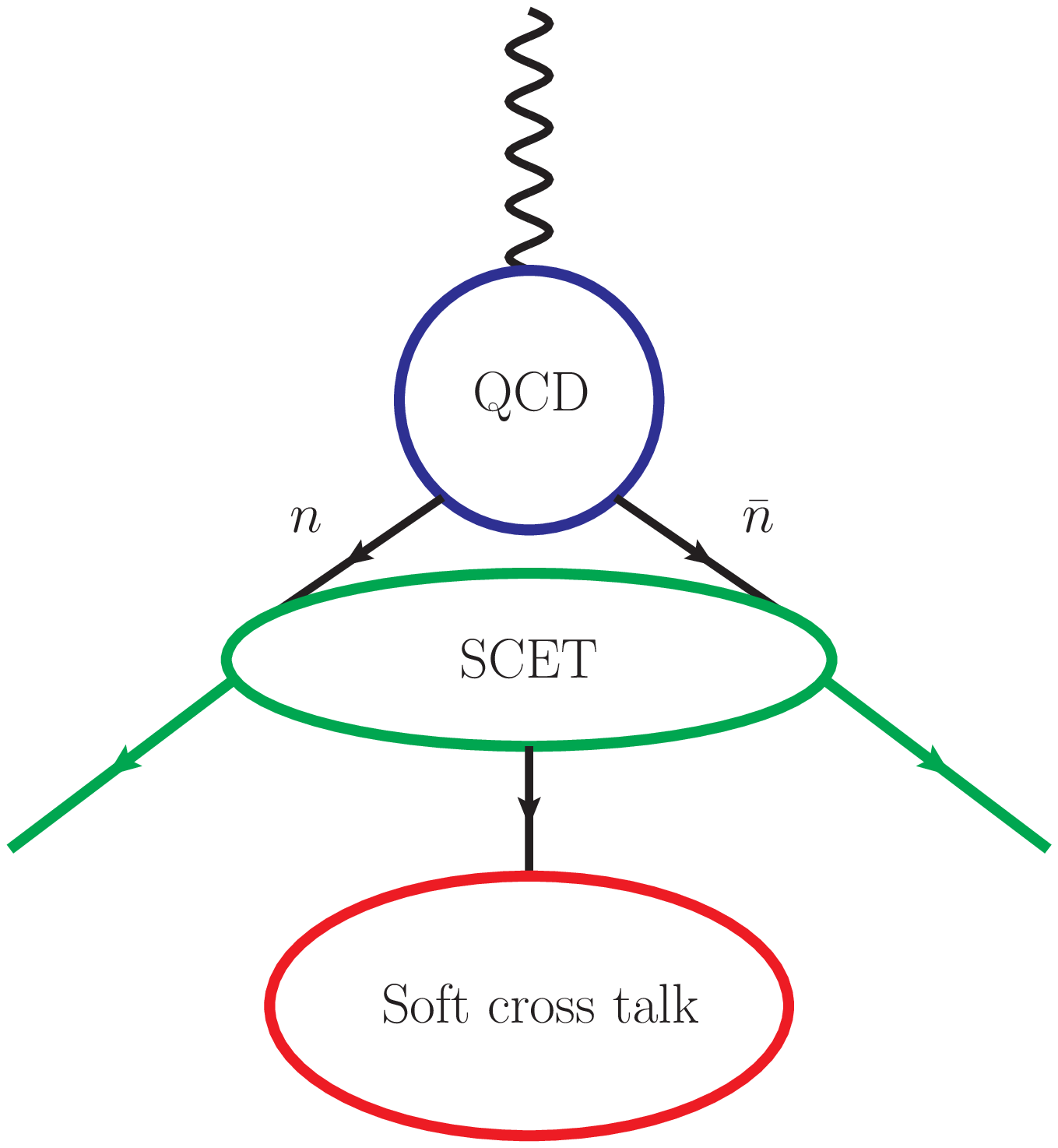}~~~~~~~~~~~
\includegraphics[width=0.3\linewidth]{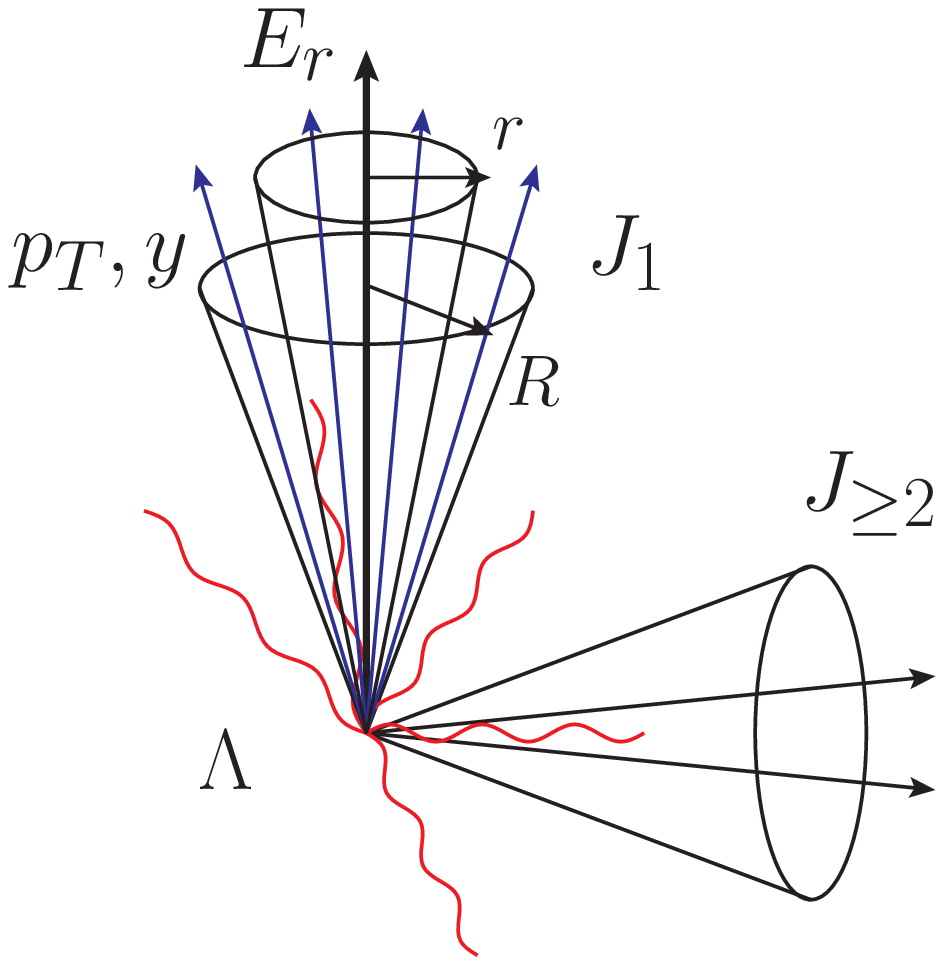}
\caption{The factorization of the hard, collinear and soft sectors in SCET (left), and the schematic event topology of $N$-jet production in $e^+e^-$ collisions (right). Notations are explained in text.}
\label{fig:SCET}
\end{figure}

SCET is an effective field theory of QCD with a systematic power counting. The key ingredient in SCET is the factorization theorem at leading power, which is the manifestation of the separation of hierarchical scales in the problem. In events with the production of energetic jets, at least three distinct scales exist: the hard scale where the hard scattering process occurs, the jet scale which describes how wide a jet is spread out in angle, and the soft scale which the energy of the rest of the radiation in the event can be (Fig. \ref{fig:SCET}). The contributions from different sectors to a physical cross section are described by the hard, jet and soft functions respectively.

Let's write down the factorization theorem of the jet shape and explain all its ingredients. At leading power, without loss of generality we can just look at the jet shape in $e^+e^-$ collisions (Fig. \ref{fig:SCET}) because at hadron colliders the contributions from initial state radiation are power suppressed \footnote{At hadron colliders, dynamical threshold enhancement also ensures that the partonic phase space with small power counting parameters dominates in the cross section calculation \cite{Appell:1988ie, Catani:1998tm, Becher:2007ty, Chien:2012ur}.}. We will motivate why the factorization theorem takes this form, and see \cite{Chien:2014nsa, Ellis:2010rwa} for detailed derivations and discussions about a related factorization theorem.  The differential cross section of $N$-jet production with the measurements of transverse momenta $p_{T_i}$ and rapidity $y_i$ of jets, an energy $E_r$ inside the cone of size $r$ in jet 1, and an energy cutoff $\Lambda$ for the radiation outside all jets can be factorized into a product of the hard, jet and soft functions.
\be
    \frac{1}{\sigma_0}\frac{d\sigma}{dE_rdp_{T_i}dy_i}
    =H(p_{T_i},y_i,\mu)J_{\omega_1}(E_r,\mu) J_{\omega_2}(\mu)\dots J_{\omega_N}(\mu)S_{n_1\dots n_N}(\Lambda,\mu)+{\cal O}(\frac{\Lambda}{Q})+{\cal O}(R).
\ee
Here, $H(p_{T_i},y_i,\mu)$ is the hard function which is square of the matching coefficient of SCET to QCD at the hard scale. $J_{\omega}(E_r,\mu)$ is the jet function which is the probability of measuring an energy $E_r$ inside a cone of size $r$ in a jet with energy $\omega=2E_J$. Here we give its operator definition,
\be
    J_{\omega}(E_r,\mu)=\sum_{X_c}\langle 0|\bar\chi_{\omega}(0)|X_c\rangle\langle X_c|\chi_{\omega}(0)|0\rangle\delta(E_r-\hat E^{<r}(X_c))\;,
\ee
and $\chi_\omega$ is the collinear jet field in SCET. All the other jet functions are unmeasured jet functions \cite{Ellis:2010rwa}. $S_{n_1\dots n_N}(\Lambda,\mu)$ is the soft function,
\be
    S_{n_1n_2\dots n_N}(\Lambda,\mu)=\sum_{X_s}\langle 0|{\cal O}_s^\dagger(0)|X_s\rangle\langle X_s|{\cal O}_s(0)|0\rangle \Theta (\Lambda-\hat E^{>R}(X_s))\;,
\ee
and ${\cal O}_s(0)$ is the product of $N$ soft Wilson lines along the jet directions. Note that the factorization theorem is a simple product instead of a convolution, especially among the collinear and the soft sectors. This is because the contribution from the soft radiation to $E_r$ is power suppressed. In general, the hard, jet and soft functions as well as their anomalous dimensions can be calculated order by order at their characteristic scales. Large logarithms of the ratio between hierarchical energy scales are resummed through the renormalization-group evolution of these functions.

Next, the {\sl averaged} energy inside the cone of size $r$ in jet 1 is
\be
    \langle E_r\rangle_{\omega_1}
    =\frac{\int dE_r E_r\frac{1}{\sigma_0}\frac{d\sigma}{dE_rdp_{T_i}dy_i}}{\frac{1}{\sigma_0}\frac{d\sigma}{dp_{T_i}dy_i}}
    =\frac{\cancel{H(p_{T_i},y_i,\mu)}J^{E_r}_{\omega_1}(\mu) \cancel{J_{\omega_2}(\mu)}\dots \cancel{S_{n_1\dots n_N}(\Lambda,\mu)}}{\cancel{H(p_{T_i},y_i,\mu)}J_{\omega_1}(\mu) \cancel{J_{\omega_2}(\mu)}\dots \cancel{S_{n_1\dots n_N}(\Lambda,\mu)}}
    =\frac{J^{E_r}_{\omega_1}(\mu)}{J_{\omega_1}(\mu)}\;,
\ee
and $J^{E_r}_{\omega}(\mu)=\int dE_r E_r~J_\omega(E_r,\mu)$ is referred to as the jet energy function. Note that most of the factors in the factorization theorem cancel out by normalizing with the differential jet rate. This implies that at leading power the jet shape is insensitive to the hard scattering process and the presence of other jets. It only depends on the energy and the partonic origin (quark or gluon) of jets. Finally, we average over the jet production cross sections with proper phase space cuts on $p_T$ and $y$,
\be
    \Psi(r)=\frac{1}{\sigma_{\rm total}}\sum_{i=q,g}\int_{PS} dp_Tdy \frac{d\sigma^{i}}{dp_Tdy}\Psi^i_\omega(r)\;,~{\rm where}~
    \Psi_\omega(r)
    =\frac{\langle E_r\rangle_{\omega}}{\langle E_R\rangle_{\omega}}
    =\frac{J^{E_r}_{\omega}(\mu)}{J^{E_R}_{\omega}(\mu)}\;.
\ee

\section{Renormalization-Group Evolution and Resummation}

\begin{figure}
\psfrag{x}{$r$}
\psfrag{z}{$\frac{d\Psi(r)}{dr}$}
\includegraphics[width=0.5\linewidth]{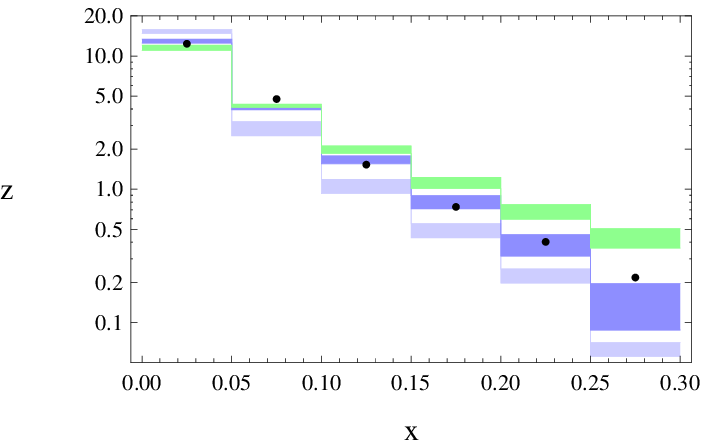}~~
\includegraphics[width=0.51\linewidth]{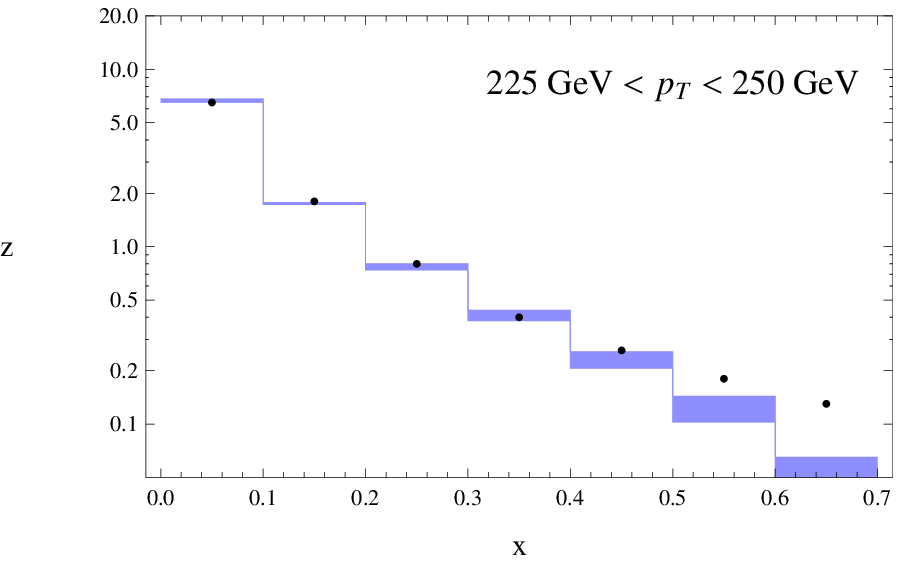}
\caption{The differential jet shape for $R=0.3$ anti-$\rm k_T$ jets with $p_T>100$ GeV and $0.3<|y|<2$ in proton-proton collisions with the center of mass energy at 2.76 TeV (left). The black dots are the CMS data. The shaded blue boxes are the LO (light) and NLL (dark) results, while the shaded green boxes are the NLL result for cone jets. As we can see, resummation and the algorithm dependence of the jet shape are important, and the data agrees with the NLL result very well. In the tail region ($r\approx R$) there is larger power corrections of ${\cal O}(r/R)$ which cause the discrepancy. The right plot is the differential jet shape for anti-$\rm k_T$ jets at the 7 TeV LHC with $R=0.7$.}
\label{fig:resum}
\end{figure}

The renormalization-group evolution of the jet energy function allows us to resum the jet shape. The RG equation is,
\be
    \frac{d J^{iE_r}_\omega(\mu)}{d\ln\mu}
    =\left[-C_i\Gamma_{\rm cusp}(\alpha_s)\ln\frac{\omega^2\tan^2\frac{R}{2}}{\mu^2}-2\gamma^{i}(\alpha_s)\right]J^{iE_r}_\omega(\mu)\;,
\ee
where we assume Casimir scaling and $\Gamma_{\rm cusp}$ is the cusp anomalous dimension. The equation can be solved and the jet energy function can be evolved from its natural scale $\mu_{j_r}$ to the renormalization scale $\mu$. The resummed integral jet shape is therefore
\be
    \Psi^i_\omega(r)
    =\frac{J^{iE_r}_\omega(\mu_{j_r})}{J^{iE_R}_\omega(\mu_{j_R})}\exp[-2C_iS(\mu_{j_r},\mu_{j_R})+2A_{i}(\mu_{j_r},\mu_{j_R})] \left(\frac{\mu^2_{j_r}}{\omega^2\tan^2\frac{R}{2}}\right)^{C_iA_\Gamma(\mu_{j_R},\mu_{j_r})}\;,
\ee
where $i=q, g$ with $C_q=C_F$ and $C_g=C_A$ the Casimir operators of the fundamental and adjoint representations in QCD. Here $S(\nu,\mu)=-\int_{\alpha_s(\nu)}^{\alpha_s(\mu)}d\alpha\frac{\Gamma_{\rm cusp}(\alpha)}{\beta(\alpha)}\int_{\alpha_s(\nu)}^\alpha\frac{d\alpha'}{\beta(\alpha')}$, $A_i(\nu,\mu)=-\int_{\alpha_s(\nu)}^{\alpha_s(\mu)}d\alpha\frac{\gamma^i(\alpha)}{\beta(\alpha)}$ and $A_\Gamma(\nu,\mu)=-\int_{\alpha_s(\nu)}^{\alpha_s(\mu)}d\alpha\frac{\Gamma_{\rm cusp}(\alpha)}{\beta(\alpha)}$ are the RG evolution kernels in SCET which resums the large logarithms. From the fixed order calculation of jet energy functions (see \cite{Chien:2014nsa}), the natural jet scale $\mu_{j_r}$ can be identified as $\omega\tan\frac{r}{2}$ which eliminates large logarithms in $J^{E_r}_\omega(\mu_{j_r})$. Thus the hierarchy between $r$ and $R$ gives two hierarchical jet scales $\mu_{j_r}$ and $\mu_{j_R}$. We compare our resummed results with the CMS measurements at the 2.76 TeV and 7 TeV LHC (Fig. \ref{fig:resum}) with good agreement. The theoretical uncertainties are estimated by varying the jet scales in the resummed expressions, shown as the shaded boxes in the plots. The results for the 2.76 TeV LHC set the baseline calculation to study the medium modification of jet shapes which we now move on to discuss.

\section{Medium Modification of the Jet Shape}

\begin{figure}
\psfrag{x}{$r$}
\psfrag{w}{\small $\frac{\rho(r)^{\rm Pb}}{\rho(r)^{\rm P}}$}
\center
\includegraphics[width=0.5\linewidth]{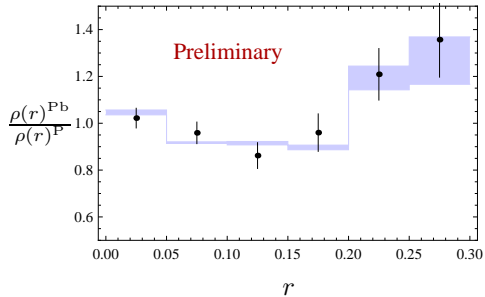}
\caption{Medium modification of jet shapes in lead-lead collisions with nucleon-nucleon center of mass energy at 2.76 TeV and centrality 0 - 10\%. The black dots are the CMS data, and the theoretical uncertainties are represented by the shaded boxes. The jet shape is sensitive to the parameters of the static QGP model. This allows us to probe the properties of the medium more precisely. The attenuation at mid $r$ and the enhancement at the periphery of the jet agree with the CMS data very well.}
\label{fig:qgp}
\end{figure}

In heavy ion collisions, jets are produced and modified as they propagate through the medium. Ref. \cite{Vitev:2008rz} calculates the medium modification of jet shapes using the Gyulassy-Levai-Vitev formalism \cite{Gyulassy:2000er,Gyulassy:2000fs} with soft gluon emissions (commonly referred to as the energy loss approach). Also see \cite{Ma:2013uqa,Ramos:2014mba} for Monte Carlo simulations of jet shapes in lead-lead collisions at the LHC. Here we build upon \cite{Chien:2014nsa} and incorporate the medium effects which have been identified as from the Glauber gluon interactions in extended SCET. Glauber gluons are off-shell modes which describe the momentum transfer transverse to the jet direction. Therefore they are not final state particles in jets which directly contribute to the measurement of jet shapes. In the framework we use, Glauber gluons are treated as background fields created from the color charges in the QGP. The dynamics of QGP is not yet calculable in first principle, and as a first attempt we use the simplest static QGP model with phenomenological parameters. With this setup, the medium induced splitting functions \cite{Ovanesyan:2011kn, Fickinger:2013xwa} have been calculated which we use in the calculation of the modification of jet energy functions. In general, the SCET jet functions can be calculated from integrating the splitting functions over appropriate phase space corresponding to the definition of jet observables. In the case of the jet energy function which plays the key role in the jet shape calculation, at leading order,
\be
    J^{iE_r}(\mu)=\sum_{j,k}\int_{PS} dxdk_\perp\frac{dN_{i\rightarrow jk}}{dxd^2k_\perp} E_r(x,k_\perp)\;,
\ee
where $\frac{dN_{i\rightarrow jk}}{dxd^2k_\perp}$ is the splitting function at ${\cal O}(\alpha_s)$. It is then straightforward to calculate the integral jet shape in heavy ion collisions. 
\be
    \Psi(r)=\frac{J^{E_r}_{v}(r)+J^{E_r}_{m}(r)}{J^{E_R}_{v}+J^{E_R}_{m}}
    =\frac{\Psi_{v}(r)J^{E_R}_{v}+J^{E_r}_{m}(r)}{J^{E_R}_{v}+J^{E_R}_{m}}\;.
\ee
Because of the Landau-Pomeranchuk-Migdal (LPM) effect, $J^{E_r}_{m}(r)$ contributes as a power correction of ${\cal O}(r/R)$ with no large logarithms at first order in opacity. To better illustrate, in the small-$x$ limit the medium induced splitting function for the $q\rightarrow qg$ channel is
\be
    \frac{dN^{m}_{q\rightarrow qg}}{dxd^2k_\perp}
    =\frac{C_F\alpha_s}{\pi^2}\frac{1}{x}\int_0^L\frac{d\Delta z}{\lambda}\int d^2q_\perp \frac{1}{\sigma_{el}}\frac{d\sigma_{el}}{d^2q_\perp}\frac{2k_\perp\cdot q_\perp}{k_\perp^2(q_\perp-k_\perp)^2}
    \Big[1-\cos\Big(\frac{(q_\perp-k_\perp)^2\Delta z}{x \omega}\Big)\Big]\;,
\ee
with the effective cross section $\frac{1}{\sigma_{el}}\frac{d\sigma_{el}}{d^2q_\perp}=\frac{m^2}{\pi(q_\perp^2+m^2)^2}$. There is no extra soft-collinear divergence when integrating over the appropriate phase space due to the oscillatory cosine term (the LPM effect), and the renormalization-group evolution of the jet energy function is the same as in vacuum. 

We calculate the jet shape for $R=0.3$ anti-$\rm k_T$ jets in lead-lead collisions with nucleon-nucleon center of mass energy at $\sqrt{s_{\rm NN}}=2.76$ TeV. The cuts $p_T>100$ GeV and $0.3<|y|<2$ are imposed. In the static QGP model we use the following typical parameters: the bulk size of QGP $L=5$ fm, the gluon mean free path $\lambda=1$ fm, and the inverse range of the glauber gluon interaction $m=0.75$ GeV. Fig. \ref{fig:qgp} shows the ratio between the jet shapes in proton-proton and lead-lead collision as a indication of the modification. Our calculation agrees with the data very well.

\section{Conclusions and Outlook}

The jet shape is resummed at NLL accuracy using the renormalization-group techniques in SCET. The baseline calculation in proton-proton collisions has been established. The LO calculation can not describe the data well and resummation is essential. The medium modification is captured by the Glauber gluon interactions which give important power corrections. We find good agreement between our calculations and the data for jet shapes in both proton-proton and lead-lead collisions. We will look into more realistic QGP models and make predictions for the upcoming 5.02 TeV run at the LHC. Hopefully better analytic understanding and precision measurements of the QGP properties will become possible in the near future.

\section{Acknowledgments}

Y.-T. Chien would like to thank the organizers of the QCD Evolution 2014 workshop for the invitation to present the work. This work is supported by the US Department of Energy, Office of Science.


\end{document}